\documentclass[prd,aps,showpacs,preprintnumbers,amssymb]{revtex4}
\usepackage{graphicx}% Include figure files

\def\e3p{$\eta \rightarrow 3 \pi$}

\begin{document}

\title{%
\hfill{\normalsize\vbox{%
\hbox{\rm SU-4252-888}
 }}\\
{Remark on pion scattering lengths}}

\author{Deirdre Black
$^{\it \bf a}$~\footnote[1]{Email:
 black@hep.phy.cam.ac.uk}}

\author{Amir H. Fariborz
$^{\it \bf b}$~\footnote[2]{Email:
   fariboa@sunyit.edu          }}

\author{Renata Jora
$^{\it \bf c}$~\footnote[3]{Email:
cjora@physics.syr.edu }}

\author{Nae Woong Park
$^{\it \bf d,e}$~\footnote[4]{Email:
 nwpark@jnu.ac.kr}}

\author{Joseph Schechter
 $^{\it \bf d}$~\footnote[5]{Email:
 schechte@phy.syr.edu}}

\author{M. Naeem Shahid
$^{\it \bf d}$~\footnote[6]{Email:
   mnshahid@phy.syr.edu            }}

\affiliation{$^ {\bf \it a} $Department of Physics,
Cavendish Laboratory, Cambridge, CB3 OHE, UK}

\affiliation{$^ {\bf \it b}$ Department of Mathematics/Science,
State University of New York Institute of Technology, Utica,
NY 13504-3050, USA.}

\affiliation{$^ {\bf \it c}$ INFN Roma, Piazzale A Moro 2,
Roma, I-00185 Italy.}

\affiliation{$^ {\bf \it d}$ Department of Physics,
 Syracuse University, Syracuse, NY 13244-1130, USA,}

\affiliation{$^ {\bf \it e}$ Department of Physics,
 Chonnam National University, Gwangju, 500-757,
 South Korea,}

\date{\today}

\begin{abstract}
   First it is shown that the tree amplitude for pion 
pion scattering in the minimal linear sigma model has
an exact expression which is proportional to a geometric
series in the quantity (s-$m_\pi^2$)/($m_B^2-m_\pi^2$), where
$m_B$ is the sigma mass which appears in the Lagrangian
and is the only a priori unknown parameter in the model.
This induces an infinite series for
 every predicted scattering length in which each 
term corresponds to a given order in the chiral perturbation
theory counting.
   It is noted that, perhaps surprisingly, the
 pattern, though not the exact values, of chiral 
perturbation
theory predictions for both the isotopic spin 0
and isotopic spin 2 s-wave pion-pion scattering lengths 
to orders $p^2$, $p^4$ and $p^6$ seems to agree with this
induced pattern. The values of the $p^8$ terms
 are also given for comparison with a possible  
future chiral perturation theory calculation.
Further aspects of this approach and future directions
are briefly discussed.
\end{abstract}


\maketitle

\section{Introduction}
    The chiral perturbation theory approach \cite{dgh}-\cite{l}
 provides
a systematic method for improving the "current algebra"
or tree level ``non-linear chiral Lagrangian" results for
 low energy QCD in powers of
a characteristic squared momentum, $p^2$ (or number of
derivatives). Intuitive understanding of the resulting
physics in some cases has been obtained by computing
the amplitudes of interest based on pole-dominance.
For example vector meson dominance is known to be 
 good at low energies; a typical well known immediate
prediction
gives the squared charge radius of the pion simply
as $r_\pi^2=6/m_\rho^2$. This kind of approach
may be theoretically justified to some extent by 
invoking the $1/N$ expansion of QCD 
\cite{tH}, \cite{W} which yields tree
level dominance.

    In the case of the pion s-wave scattering
 lengths, the long controversial, but now
 apparently accepted, sigma
particle would appear to play the role of the
 rho meson. However, a simple sigma dominance
approximation is not viable because it would
not guarantee the nearly spontaneous breakdown
of chiral symmetry mechanism which is crucial
for QCD. Such a mechanism is guaranteed by the
use of a linear sigma model of some type. 
The properties
of the sigma suggest that
 it may be a four quark
(i.e. $qq{\bar q}{\bar q}$) state of some kind or a 
mixture of four quark and two quark
 components \cite{fjs}. In such
an instance the $1/N$ expansion
 would not hold in
 its usual form \cite{ss} and those models have a
 lot of subtleties.
Here, we will not discuss such questions but
will just point out that the minimal 
SU(2) linear
sigma model \cite{gl}
provides a useful approximation to the lightest
sigma of a model which may contain a number of them.
A crucial effect is that the linear sigma model 
has an important contact term. The actual low
energy scattering is known to result from an enormous 
cancellation between the sigma pole and the contact
contributions. This unpleasant feature is mitigated
in the non-linear sigma model (which forms the basis
of the chiral perturbation scheme). Another way to mitigate
this feature is at the amplitude level. Then the 
amplitude is expanded \cite{fjs1}, \cite{fjs2}
 in a Taylor series 
about $s=m_\pi^2$ and the cancellation may be explicitly
made. The result is proportional to a simple geometric
series in the variable $(s-m_\pi^2)/(m_B^2-m_\pi^2)$.
Then in order to compare it with something, it is
natural to compare it with another power series in 
 squared momentum- chiral perturbation theory.

\section{Expanded scattering lengths}

   With the Mandelstam notation, the invariant
pion scattering amplitude computed at tree
 level in the minimal
SU(2) linear sigma model reads:
\begin{equation}
A(s,t,u)= \frac{2(m_B^2-m_\pi^2)}{F_\pi^2}
        \left[(1-\frac{s-m_\pi^2}{m_B^2-m_\pi^2})^{-1}
          -1\right],
\label{stuamp}
\end{equation}
where $F_\pi$= 131 MeV and $m_B$ denotes the ``bare"
sigma mass which appears in the Lagrangian.

This equation is seen to contain a contact term as well
as a pole term which has been rewritten for convenience.
In this form it is apparent that there is a
 geometric series expansion in powers of
 $(s-m_\pi^2)/(m_B^2-m_\pi^2)$, which should be rapidly
convergent for $s$ close to the pion- pion
threshold:

\begin{equation}
A(s,t,u)= \frac{2(s-m_\pi^2)}{F_\pi^2}
        \left[1
         +\frac{s-m_\pi^2}{m_B^2-m_\pi^2}
          +\frac{(s-m_\pi^2)^2}{(m_B^2-m_\pi^2)^2}
         +\frac{(s-m_\pi^2)^3}{(m_B^2-m_\pi^2)^3}
          +\cdots
          \right].
\label{2stuamp}
\end{equation}
 Actually, a similar expansion may be
derived when a number of different scalar mesons
are present \cite{fjs2}.In that instance the 
lowest
lying scalar meson is expected to dominate near threshold. 

    The isospin 0 scattering length is proportional
to $3A(s,t,u)+A(t,s,u)+A(u,t,s)$ evaluated at
$s=4m_\pi^2,t=u=0$ while the isospin 2 scattering 
length is obtained by evaluating $A(t,s,u)+A(u,t,s)$
instead. Then we find for the ``dimensionless"
s-wave scattering lengths:
\begin{equation}
m_\pi a_0^0 = \frac{m_\pi^2}{16\pi F_\pi^2}\left[
              7 +29\frac{m_\pi^2}{m_B^2-m_\pi^2}
                +79\frac{m_\pi^4}{(m_B^2-m_\pi^2)^2}
                +245\frac{m_\pi^6}{(m_B^2-m_\pi^2)^3}
                +\cdots \right],
\label{expand0}
\end{equation}

and, 
\begin{equation}
m_\pi a_0^2 = -\frac{m_\pi^2}{8\pi F_\pi^2}\left[
              1 -\frac{m_\pi^2}{m_B^2-m_\pi^2}
                +\frac{m_\pi^4}{(m_B^2-m_\pi^2)^2}
                -\frac{m_\pi^6}{(m_B^2-m_\pi^2)^3}
                +\cdots \right].
\label{expand2}
\end{equation}
Evidently, these terms may be consecutively
 interpreted as $p^2$, 
$p^4$, $p^6$, and $p^8$ etc. contributions.

\section{Numerical comparison}
The chiral perturbation theory results to the
first three orders \cite{w}, \cite{gal},\cite{cgl}
 as well as the comparison with 
experiment may be conveniently read from Fig. 10
of \cite{pislak}. We have subtracted the values
presented there to get the incremental corrections
for comparison with Eqs.(\ref{expand0}) and
 (\ref{expand2}). The order $p^2$ entries \cite{w}
in Table \ref{firsttable} are of
 course the same and we made the choice $m_\pi$ =
 140 MeV to enforce this feature. The only unfixed
parameter in the linear sigma model is the bare 
sigma mass, $m_B$ which we chose to be 550 MeV
to give a $p^4$ contribution to the resonant 
partial wave scattering length which approximately agrees
 with chiral perturbation theory at that order.
(Alternatively, a similar value can be found on 
an a priori basis by fitting the near
threshold I=0, s-wave scattering data).
\begin{table}[htbp]
\begin{center}
\begin{tabular}{c|c|c|c|c}
\hline
  Order: & $p^2$ & $p^4$ & $p^6$ & $p^8$  \\
\hline \hline
$m_{\pi} a_0^0$ in ChPT: & 0.16 & 0.04 & 0.02 $\pm$ 0.005 & -      \\
$m_{\pi} a_0^0$ in LSM: &  0.159 &  0.046 & 0.009 & 0.0019   \\
\hline
$m_{\pi} a_0^2$ in ChPT: & -0.046 & 0.004 & -0.002 $\pm$ 0.001
& -        
\\
$m_{\pi} a_0^2$ in LSM: &  -0.0454 & 0.0031  & -0.0002 & 0.000015       \\
\hline
\end{tabular}
\end{center}
\caption[]{Comparison of scattering length increments
}
 \label{firsttable}
\end{table}

    We notice that the increments to $m_{\pi} a_0^2$
are predicted to alternate in sign with increasing order.
This pattern manifestly agrees with what was found in the
first three orders of chiral perturbation theory.

    If the $p^4$ increment of  $m_{\pi} a_0^0$ is taken
 as approximately a common input, the magnitude of
the $p^4$ increment to $m_{\pi}
 a_0^2$ is predicted to be about 75 percent of the chiral
perturbation theory one. Also the magnitude of 
the $p^6$ increment to $m_{\pi} a_0^0$ is predicted to
be about 50 percent of the chiral perturbation theory one.
Finally, the magnitude of the $p^6$ increment
 to $m_{\pi} a_0^0$  could be about 20 percent of 
the chiral perturbation theory one (which contains a 
large uncertainty however). Thus it seems fair to say that
the tree level linear sigma model result exactly reproduces
the signs of the chiral perturbation amplitudes and tracks
well the magnitudes. It will be interesting to compare
the predicted $p^8$ increments given above when the chiral 
perturbation theory calculation is carried to that order.

   Differences between the chiral perturbation results
for the s-wave scattering lengths
 and the present ones may be evidently
 interpreted physically as 
due to contributions from effects other than the
existence of the sigma meson.
It is likely that the next most important effects
should arise from including the rho meson and a higher
mass scalar meson like the $f_0(980)$in the
formulation of the chiral invariant linear sigma model.
Work in this direction is under way.

\section{Discussion}

    Different recent discussions of the pion scattering
lengths in the linear sigma model are given in 
\cite{skr}, \cite{a}, \cite{pgr}.

    The main new feature in the present approach seems to be
the realization that the use of the simplest linear sigma model
at tree level
does not give just one number (a scattering length)
but gives an infinite series of numbers which
can be conveniently compared with the series resulting
from chiral perturbation theory.

    Another amusing feature is that this approach
 provides
a specific model for the expansion parameter of this
series; namely $m_\pi^2/(m_B^2-m_\pi^2)$.

   Of course, in comparison
 with chiral perturbation theory,
there is an obvious difference in that the latter
 approach includes the effect of loop integrals.
 The loop integrals
enforce that chiral perturbation theory
carried to all orders should result in fully
unitarized scattering amplitudes. In the
 present approach it is possible to obtain exactly
 unitary partial wave
 amplitudes without
 introducing any new parameters by means of the
K-matrix technique. For example, in a  
3 flavor linear sigma model \cite{unitarized},
the resonant partial wave
amplitude up to about 1.2 GeV was simply
fit by including 
just the
sigma and $f_0(980)$ scalar mesons
together with K-matrix unitarization.
An approach equivalent
to K matrix unitarization for the 2 flavor linear
sigma model was described in \cite{as}.
 It is easy to see \cite{nochange} that 
K matrix unitarization actually does not change the
predicted value of the scattering length from the one
determined at tree level.

     It is well known that to
accurately model low energy pion
 physics it is necessary
 to take
the rho meson into account in addition
 to the sigma. So certainly, the next step
in the present approach is to investigate
the power series expansion of the 
pion scattering amplitude computed from a
 linear sigma model in
 which the rho meson as well as
an axial vector meson (for chiral symmetry)
are included. This can be expected to improve
the agreement with the chiral perturbation
theory expansion. (Of course the pion scattering has been
treated from other points of view in
 {\it non-linear} models
containing both scalars and vectors; see for examples,
\cite{hss}-\cite{hss2}).

\section*{Acknowledgments} \vskip -.5cm
We are happy to thank
 A. Abdel-Rehim, M. Harada,
S. Moussa, S. Nasri and F. Sannino
 for helpful related discussions.
The work of D.B. is supported by a Royal Society
 Dorothy Hodgkin Fellowship.
The work of A.H.F. has been partially
supported by the NSF Grant 0652853.
NWP is 
supported by Chonnam National University.
The work of
 J.S. and M.N.S. is supported in part by the U.
S. DOE under Contract no. DE-FG-02-85ER 40231.


\begin{thebibliography}{10}

\bibitem{dgh} References and discussion of
 the connection with vector meson dominance are
given in J.F. Donoghue, E. Golowich and 
B.R. Holstein, ``Dynamics of the Standard Model",
Cambridge University Press, 1992.

\bibitem{egpd} G. Ecker, J. Gasser, A. Pich
and E. de Rafael, Nucl. Phys {\bf B321}, 311(1999).

\bibitem{oo}J.A. Oller and E. Oset,
 Nucl. Phys. {\bf A620}, 
438 (1997).

\bibitem{l}H. Leutwyler, arXiv:hep-ph/0608218. 

\bibitem{tH}G.'t Hooft, Nucl. 
Phys. {\bf B72}, 461(1974).

\bibitem{W}E. Witten, Nucl. Phys. {\bf B160}, 57 
(1979).

\bibitem{fjs}Many references are given in
A.H. Fariborz, R. Jora and J. Schechter,
arXiv:0902.2825 [hep-ph].

\bibitem{ss}See for example,
F. Sannino and J. Schechter, Phys. Rev. D 
{\bf 76}, 014014 (2007). 

\bibitem{gl}By the minimal model we mean
just the meson terms in M. Gell-Mann and
M. Levy, Nuovo Cimento {\bf 16}, 705 (1960).

\bibitem{fjs1}See Eq.(38)of
 A.H. Fariborz, R. Jora and J. Schechter,
Phys. Rev. D {\bf 76}, 114001 (2007).  
 In this model there
are four different scalars.

\bibitem{fjs2}See Eq.(68) of  A.H. Fariborz,
 R. Jora and J. Schechter, Phys. Rev. D
{\bf 77},034006 (2008).

\bibitem{w}S. Weinberg, Phys. Rev. Lett.{\bf 17},
616 (1966).

\bibitem{gal}The order $p^4$ calculation is from
J. Gasser and H. Leutwyler, Phys. Lett. {\bf 125B},
325 (1983).

\bibitem{cgl}The order $p^6$ calculation is from
G. Colangelo, J. Gasser and H. Leutwyler,
Phys. Lett. {\bf 488B}, 261
(2000).

\bibitem{pislak}S.Pislak et al, Phys. Rev.D {\bf 67},
072004 (2003). 

\bibitem{skr}M.D. Scadron, 
F. Kleefeld and G. Rupp,
arXiv:hep-ph/0601196.

\bibitem{a}N. Achasov,
 arXiv:0810.2201[hep-ph].

\bibitem{pgr}D. Parganlija, F. Giacosa and D.H. Rischke,
arXiv:0812.2183[hep-ph].

\bibitem{unitarized}D. Black, A.H. Fariborz,
S. Moussa, S. Nasri and J. Schechter, Phys. Rev.
D {\bf 64}, 014031 (2001).

\bibitem{as}An approach equivalent
to K matrix unitarization for the 2 flavor linear
sigma model was described in N.N. Achasov and 
G.N. Shestakov, Phys. Rev. D {\bf 49}, 1994. 

\bibitem{nochange}See the discussion around Eq.(47)
in \cite{fjs1} given above.

% \bibitem{different}
% See Figures 3 and 4 of 
% A. Abdel-Rehim,
%  D. Black, A.H. Fariborz,
% S. Nasri and J. Schechter Phys. Rev. D {\bf 
% 68}, 013008 (2003).

\bibitem{hss}M. Harada, F. Sannino and 
J.Schechter, Phs. Rev. D {\bf 54}, 1991
 (1996).

\bibitem{oop}J.A. Oller, E. Oset 
and J.R. Pelaez, Phys. Rev. Lett. {\bf 80},
3452 (1998).

\bibitem{ih}K. Igi and K. Hikasa, Phys. Rev.
 D {\bf 59}, 034005 (1999).

\bibitem{hss2}M. Harada, F. Sannino
 and J. Schechter, Phys. Rev. Lett.
 {\bf 78}, 1603 (1997).

\end{thebibliography}
\end{document}